\documentclass[10pt,conference]{IEEEtran}
\IEEEoverridecommandlockouts
\usepackage{cite}
\usepackage{amsmath,amssymb,amsfonts}
\usepackage{algorithmic}
\usepackage{graphicx}
\usepackage{textcomp}
\usepackage{xcolor}
\usepackage{makecell}
\usepackage{booktabs}

\def\BibTeX{{\rm B\kern-.05em{\sc i\kern-.025em b}\kern-.08em
    T\kern-.1667em\lower.7ex\hbox{E}\kern-.125emX}}
\begin{document}

\title{Do code refactorings influence the merge effort?\\
\thanks{The authors would like to thank CNPq (grants 311955/2020-7, 141054/2019-0, and 315750/2021-9), FAPERJ (grants E-26/010.101250/2018, E-26/010.002285/2019, E-26/211.033/2019, E-26/201.038/2021, and E-26/201.139/2022), and IEEA-RJ (grant 001/2021) for the financial support.}
}
\author{\IEEEauthorblockN{André Oliveira\IEEEauthorrefmark{1}, Vânia Neves\IEEEauthorrefmark{1}, Alexandre Plastino\IEEEauthorrefmark{1}, Ana Carla Bibiano\IEEEauthorrefmark{2}, Alessandro Garcia\IEEEauthorrefmark{2}, and Leonardo Murta\IEEEauthorrefmark{1}}
\IEEEauthorblockA{\IEEEauthorrefmark{1}Instituto de Computação (IC), Universidade Federal Fluminense (UFF), Niterói, Brazil 24210--346\\Email: andrelucio@id.uff.br, \{vania, plastino, leomurta\}@ic.uff.br}
\IEEEauthorblockA{\IEEEauthorrefmark{2}Informatics Department – PUC-Rio, Rio de Janeiro, Brazil 22451--900\\Email: \{abibiano, afgarcia\}@inf.puc-rio.br}}

\maketitle

\begin{abstract}
In collaborative software development, multiple contributors frequently change the source code in parallel to implement new features, fix bugs, refactor existing code, and make other changes. These simultaneous changes need to be merged into the same version of the source code. However, the merge operation can fail, and developer intervention is required to resolve the conflicts. Studies in the literature show that 10 to 20 percent of all merge attempts result in conflicts, which require the manual developer’s intervention to complete the process. In this paper, we concern about a specific type of change that affects the structure of the source code and has the potential to increase the merge effort: code refactorings. We analyze the relationship between the occurrence of refactorings and the merge effort. To do so, we applied a data mining technique called association rule extraction to find patterns of behavior that allow us to analyze the influence of refactorings on the merge effort. Our experiments extracted association rules from 40,248 merge commits that occurred in 28 popular open-source projects. The results indicate that: (i) the occurrence of refactorings increases the chances of having merge effort; (ii) the more refactorings, the greater the chances of effort; (iii) the more refactorings, the greater the effort; and (iv) parallel refactorings increase even more the chances of having effort, as well as the intensity of it. The results obtained may suggest behavioral changes in the way refactorings are implemented by developer teams. In addition, they can indicate possible ways to improve tools that support code merging and those that recommend refactorings, considering the number of refactorings and merge effort attributes.
\end{abstract}

\begin{IEEEkeywords}
Software Merge, Merge Effort, Refactoring, Association Rules, Data Mining.
\end{IEEEkeywords}

\section{Introduction}
\label{sec:introduction}

Developers frequently change in parallel the same source code during the software development process due to time to market. Eventually, these parallel changes need to be merged. Previous work reported that 10 to 20 percent of all merges fail~\cite{brun2011, kasi2013}, some projects experiencing rates of almost 50 percent~\cite{brun2011, zimmermann2007}. The effort for merging parallel changes might be high due to various factors, such as the need to resolve conflicts. Over the years, many merge conflict resolution techniques have been developed, such as those described by Mens~\cite{mens2002} and Apel et al.~\cite{Apel2011}. These techniques differ considerably when comparing two artifact versions and how they resolve merge conflicts. There are many proposals for approaches that seek to resolve these conflicts in an automated or semi-automated way ~\cite{apiwattanapong2007, binkley1995, buffenbarger1995, hunt2002, shen2004, shen2005, westfechtel1991, berzins1994, jackson1994, lebenich2015, apel2012, Apel2011}. Despite this, developers often need to intervene in conflicts that cannot be resolved automatically, demanding manual effort. 

Code refactoring is a widely used practice to improve software modularity and, as such, facilitate parallel code changes~\cite{kim2012}. Refactoring is a code change to modify software’s internal structure, in which the resulting code is expected to bring various benefits in the long term~\cite{fowler2018}. However, one should understand the short-term effort implications of applying code refactoring in collaborative software development. Otherwise, developers may blindly apply code refactoring that increase merge effort; for instance, they might perform refactorings that lead to conflicts later, potentially increasing merge effort.

Some pieces of work in the literature~\cite{dig2007, lebenich2017, mahmoudi2018, laszlo2007, mahmoudi2019} already presented studies analyzing the effects of refactorings and code merging. Some of them~\cite{dig2007, laszlo2007, lebenich2017} have proposed tools capable of identifying a small subset of refactorings before performing the code merge. The purpose of these studies is to identify the most common refactorings, such as renaming and moving code, to assist developers in making decisions before performing the code merge. Mahmoudi and Nadi~\cite{mahmoudi2018} investigated and reported the most common types of refactorings that occur in practice and analyzed the possibility of having automated support for merging them but they have not developed corresponding tools. Mahmoudi et al.~\cite{mahmoudi2019} carried out an empirical study to assess the relationship between 15 types of refactorings and the occurrence of merge conflicts. As a result, they found that 22\% of merge conflicts involve refactorings and concluded that these conflicts are more complex than those without refactorings. Moreover, they concluded that 11\% of conflicting regions have at least one refactoring involved. 

Nevertheless, these studies have not investigated the relationship between the occurrence of refactorings and the practical effort to perform the merge operation. Moreover, studies that quantify the intensity of this relationship have not been carried out either. Another aspect that has not been analyzed concerns where refactorings are implemented in the branches of a merge commit. For example, whether their occurrence simultaneously in two branches can generate more or less merge effort. In addition, they only considered a limited subset of refactorings types. Fowler’s catalog~\cite{fowler2018} describes an expressive set of different kinds of refactorings.

Our research analyzes, from a different perspective: i) the relationship between the occurrence of refactorings in the branches of a merge commit, and ii) the effort to merge the branches. We focus on evaluating the effort required for performing the merge operation instead of analyzing the areas of conflict involving refactorings and their size, as proposed by Mahmoudi et al.~\cite{mahmoudi2019}. We adopted a descriptive data mining technique called association rule extraction to understand and quantify this relationship. The application of this technique had the purpose of analyzing how much the presence of refactorings influences the chances and the intensity of effort during the merge. 

Our experiments were carried out with data collected from 28 open-source projects hosted on GitHub and considering 33 different types of refactoring. More specifically, we seek to answer the following research questions, detailed in Section~\ref{sec:results}:

\textbf{RQ1}: Does the occurrence of refactorings in the branches increase the chances of merge effort? 

\textbf{RQ2}: Does the amount of refactorings in the branches increase the chances of merge effort? 

\textbf{RQ3}: Does the amount of refactorings in the branches increase the intensity of merge effort?

 When evaluating the merge of branches, most papers in the literature indicate when the merge fails or not or, at most, count the number of conflicting chunks. However, the main issue regarding merging is not whether they fail or have many conflicts but how hard it is to fix them. In this paper, we consider the effort performed during merge resolution. To do so, we adopted code churn as a surrogate for the merge effort. According to the literature~\cite{yamashita2013}, code churn may be a reasonable alternative for code maintenance effort, with a significant moderate to strong Spearman correlation of 0.59 to 0.66 between both metrics for corrective and evolutive maintenance, respectively.

During the research, we identified that 7.1\% of the merge commits failed; that is, it took some effort to resolve them. We emphasize that this percentage is comparable with the literature (from 10\% to 20\%, as previously mentioned). This range may appear small at first glance, but merge is a frequent operation (e.g., in our dataset, 1 in every 10.6 commits is a merge commit), and in some projects of our sample, merge fails multiple times every month. Moreover, some of these merge failures are cumbersome for the developers, negatively impacting their productivity and potentially affecting the software quality~\cite{mcKee2017, brindescu2018}.

We found that the occurrence of refactorings can increase by 24\% the chances of merge effort. Moreover, as the number of refactorings in the branches of a merge commit increases, the chances of merge effort also increase. For instance, for merges with hundreds or more refactorings, the chances of having effort increase in 143\%. In addition, we identified that the number of refactorings in the branches of a merge also influences the intensity of merge effort. We noticed that, for merges with hundreds or more refactorings, the chances of having hundreds or more lines of code changed during merge increased by 232\%. Furthermore, these percentages tend to be even higher when refactorings co-occur in both branches of the merge commit. In this case, the chances of having effort increase by 114\%. Likewise, when there are many refactorings (hundreds or more) in both branches, this percentage of having effort is even higher: 308\%. In this scenario with many refactorings, we also observed a 751\% increase in the chances of the effort intensity being high.

Our results may indicate to systems managers and developers the application of alternative ways to use branches to implement refactorings. Applying these alternatives aims to avoid or minimize the implementation of refactorings in parallel. From the tool builders' perspective, our findings may indicate the need for new merge strategies, considering the number of refactorings implemented in both branches and the predicted merge effort level. This approach could signal the most appropriate moment to perform the merge regarding a specific threshold of refactorings that may decrease the chances of high merge effort. Furthermore, our results may encourage more particular studies on which types of refactorings tend to generate more merge effort. We will discuss these possible implications in more detail in Section~\ref{sec:resultsDiscussion}.

The remainder of this paper is organized as follows. Section~\ref{sec:methodology} presents the research process adopted in our work, the dataset used in the experiments, and an overview of the employed techniques. Then, Section~\ref{sec:results} shows the obtained results and discusses their implications, answering the research questions. Section~\ref{sec:threats} discusses the threats to the validity of this study. In Section~\ref{sec:related}, we present a discussion about related pieces of work. Finally, Section~\ref{sec:conclusions} concludes this work by highlighting our main contributions and presenting perspectives for future work.

\section{Materials and Methods} \label{sec:methodology}

\begin{figure}[!t]
\centering
\includegraphics[width=2.1in]{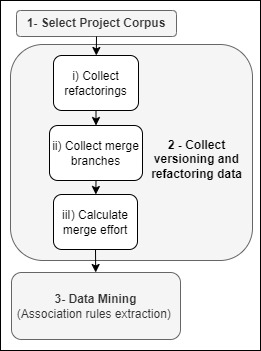}
\caption{The phases of the experiments.}
\label{fig:phases_steps}
\end{figure}

The experimental process of our work can be divided into three phases, as shown in Figure~\ref{fig:phases_steps}. In the first phase, we defined the criteria for selecting the projects analyzed in our experiments, as detailed in Section~\ref{sec:project_corpus}. The second phase aims to collect information about merge branches and refactorings, and compute the merge effort. This phase is described in Section~ \ref{sec:refac_merge_effort}. In the last phase, presented in Section~\ref{sec:association_rules}, we employed a data mining technique named association rules extraction to answer the research questions raised in Section~\ref{sec:introduction}. Applying this technique, we can discover hidden information in the analyzed data that could have been ignored if a manual exploratory analysis had been conducted. The complete experimental package, containing the data and scripts used in this research, and the reproducibility instructions, is available at https://github.com/gems-uff/refactoring-merge.

\subsection{\textbf{Project Corpus}} 
\label{sec:project_corpus}

When selecting the corpus, we aimed at mature and relevant open-source projects hosted on GitHub. We first used the GitHub GraphQL API (v4)\footnote{https://docs.github.com/pt/graphql} to search for all public repositories that were not forks of other repositories, had at least 5,000 stars, were not archived, and received at least one push in the last three months. According to Kalliamvakou et al.~\cite{kalliamvakou2014}, avoiding forks is essential to guarantee that the corpus contains only one repository per project. However, we are aware of forks that are much more successful than the original forked projects. To ensure that we were not excluding one of these projects, we checked for forked projects that met our selection criteria and found none. Moreover, restricting the number of stars to at least 5,000 guarantees that our corpus contains just relevant and popular repositories~\cite{borges2018}. Finally, avoiding archived repositories or repositories that did not receive pushes in the last three months ensures a certain degree of activity in all repositories of our corpus. This search was performed on September 20, 2021, and returned 3,201 repositories.

Afterward, we analyzed the metadata of these 3,201 repositories to perform additional filters on the number of contributors (10 or more) and the number of commits (5,000 or more) in the default branch. Filtering out repositories with less than ten contributors aims at avoiding personal or coursework projects in our corpus~\cite{kalliamvakou2014}. Moreover, restricting the number of commits in the default branch to 5,000 or more is an attempt to remove immature or short-term projects from our corpus. After applying the filter for the number of contributors, 2,941 projects remained, and of these, only 477 repositories had at least 5,000 commits. 

From these 477 repositories, we applied another filter, considering only projects whose primary language reported on GitHub is Java. The Java language was chosen considering the following criteria: i) the need to focus on a specific language, as refactoring analysis is language-dependent; ii) its popularity, according to TIOBE ranking~\cite{tiobe2022} and StackOverflow Survey~\cite{stackoverflow2021}; iii) and the fact that the tool that identifies refactorings (RefactoringMiner~\cite{tsantalis2020}), used in this work, can analyze only code written in Java. After applying this filter, 42 projects remained~\cite{tsantalis2018}.

We then conducted a manual inspection of the remaining 42 repositories, examining the GitHub repository and the web page of each project. This analysis aims to eliminate those not containing a software project and those not documented in the English language. The first criterion refined the automatic filter for the primary programming language. Even though GitHub was able to detect a primary language for all repositories in our corpus at this point, one of them stored only documentation (e.g., books, software documentation, main pages, etc.), eventually having source code used as examples (this explains the GitHub classification of primary language). Only one repository did not meet this criterion, resulting in 41. We only selected project artifacts in English to guarantee that we would be able to understand the documentation of the projects in the corpus. At this point, two projects were removed because their documentation was written in Chinese, lasting 39 projects. Even though automatic translators are available, we decided not to include those repositories in the corpus due to the poor quality of the translation for some repositories. Thus, including them would be a threat to our analysis.

\begin{figure}[!t]
\centering
\includegraphics[width=3.49in]{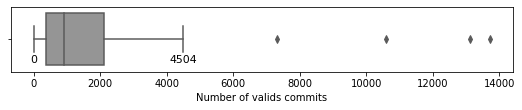}
\caption{Boxplot with the distribution of valid merge commits across projects}
\label{fig:boxplot}
\end{figure}

The last filter considered the number of valid merge commits. In our study, a valid merge commit cannot be a no-fast-forward type, as our goal is to evaluate the merge effort. A fast-forward merge can be performed when there is a direct linear path from the source branch to the destination branch. In a fast-forward merge, Git moves the target branch pointer to the exact location as the source branch pointer without creating an extra merge commit. The \textit{git merge --no-ff} command merges the specified branch into the current branch, creating a new merge commit and keeping the structure and history of the branches intact, even if a fast-forward merge is possible. We computed the distribution of valid merge commits for each project, as shown in Figure~\ref{fig:boxplot}. Considering the low and upper limits and the quartile values of this boxplot, calculated using Tukey’s fences formula~\textit{Q3 + 1.5 × IQR}~\cite{barnett1994}, we defined some other criteria to select the projects.

At first, we discarded all projects with valid merge commits above the maximum threshold (4,504 merge commits). This action was taken so that the experiments could be conducted in a more homogeneous dataset, avoiding, for example, that a large project alone dominates the overall results. After applying this filter, four projects were discarded (graal, spring-boot, neo4j, and intellij-community), resulting in 35 projects. We also removed projects with a number of valid merge commits smaller than the limit of the first quartile of the boxplot (173 merge commits). Likewise, we discarded these projects as they would have little representation in the general context of building our dataset. Seven projects did not meet this criterion (CoreNLP, bazel, buck, guava, litho, presto, and selenium), and finally, our project corpus was composed of 28 projects.

Table~\ref{table:projectcorpus} shows the characteristics of the projects selected for analysis, presenting the number of commits~(NC), the number of merge commits~(NMC), the number of merge commits using the \textit{--no-ff} flag~(NMC-nff), and the number of valid merge commits~(NVMC). The final version of the dataset discarded about 26.88\% of the merges because they were generated by the Git command \textit{--no-ff}, resulting in 40,248 merge commits.

\begin{table}[htpb]
\centering
\caption{Characteristics of the analyzed projects.}
\label{table:projectcorpus}
\begin{tabular}{lrrrrr}
\toprule[1.0pt]
\textbf{Project} &
\textbf{\begin{tabular}[c]{@{}c@{}}NC\end{tabular}} & 
\textbf{\begin{tabular}[c]{@{}c@{}}NMC\end{tabular}} & 
\textbf{\begin{tabular}[c]{@{}c@{}}NMC-nff\end{tabular}} & 
\textbf{\begin{tabular}[c]{@{}c@{}}NVMC\end{tabular}} \\

\midrule
Activiti & 10,752 & 1,917 & 682 & 1,235 \\
antlr4 & 7,966 & 1,652 & 632 & 1,020 \\
Arduino & 7,300 & 819 & 170 & 649 \\
cas & 19,884 & 5,302 & 798 & 4,504 \\
che & 9,119 & 1,238 & 626 & 612 \\
closure-compiler & 17,372 & 389 & 100 & 289 \\
dbeaver & 20,819 & 4,553 & 557 & 3,996 \\
dropwizard & 5,775 & 1,127 & 563 & 564 \\
druid & 6,413 & 1,491 & 296 & 1195 \\
elasticsearch & 60,687 & 5,123 & 1,317 & 3,806 \\
ExoPlayer & 10,751 & 597 & 155 & 442 \\
flink & 26,869 & 653 & 29 & 624 \\
gocd & 12,501 & 2,437 & 1,076 & 1,361 \\
hadoop & 25,269 & 663 & 16 & 647 \\
incubator-druid & 11,211 & 2,176 & 612 & 1,564 \\
incubator-shardingsphere & 30,067 & 2,763 & 646 & 2,117 \\
jenkins & 31,502 & 4,944 & 599 & 4,345 \\
libgdx & 14,666 & 2,677 & 584 & 2,093 \\
mockito & 5,516 & 500 & 240 & 260 \\
netty & 10,324 & 309 & 136 & 173 \\
pinpoint & 12,004 & 1,968 & 999 & 969 \\
processing & 13,198 & 1,133 & 187 & 946 \\
realm-java & 8,661 & 3,441 & 833 & 2,608 \\
redisson & 6,473 & 1,050 & 143 & 907 \\
RxJava & 5,870 & 1,566 & 678 & 888 \\
skywalking & 6,407 & 1,000 & 382 & 618 \\
spring-framework & 22,859 & 1,257 & 718 & 539 \\
zaproxy & 7,874 & 2,299 & 1,022 & 1,277 \\ 
\midrule
Total & 428,109 & 55,044 & 14,796 & 40,248 \\
\bottomrule[1.0pt]
\end{tabular}
\end{table}
  
\subsection{\textbf{Refactorings and Merge Effort}} \label{sec:refac_merge_effort}

As shown in Figure~\ref{fig:phases_steps}, the second phase was performed in 3 steps: (i) identify the merge commits that have code refactorings, (ii) collect and store the structure branches of the merge commits, indicating the commits containing refactorings in each branch, and (iii) calculate the merge effort. 




We have considered in our study 33 different types of refactorings, 26 of which are described in Fowler's catalog~\cite{fowler2018}:
Change Return Type, Extract Attribute, Extract Class, Extract Interface, Extract Method, Extract Subclass, Extract Superclass, Extract Variable, Inline Method, Inline Variable, Merge Attribute, Move Attribute, Move Class, Move Method, Pull Up Attribute, Pull Up Method, Push Down Attribute, Push Down Method,
Rename Attribute, Rename Class, Rename Method, Rename Parameter, Rename Variable, Split Attribute, Split Parameter, and Split Variable. The other seven types of refactorings were defined by Tsantalis et al.~\cite{tsantalis2020}: Change Parameter Type, Change Variable Type, Merge Parameter, Merge Variable, Parameterize Variable, Replace Attribute, and Replace Variable with Attribute. We consider these subsets of refactorings, as a recent study revealed that developers often applied them in practice~\cite{bibiano2021}.

To investigate the effects of refactorings on the merge effort, we adopted a metric defined by Prud\^encio et al.~\cite{prudencio2012} and implemented by Moura and Murta~\cite{moura2018}. A more technical explanation of how we compute the merge effort can be found in Moura and Murta~\cite{moura2018}, but we summarize it as follows. First, we identified the code churn of each branch by performing a \textit{diff} between the base version (i.e., the common ancestor) and the tip of the branch. The two sets of actions (lines of code added and removed in the branches) are combined, producing a multiset~\cite{knuth1997} with all actions performed in the branches.



Then, we identified the code churn of the merge by performing a \textit{diff} between the base version and the merge version. This produces a multiset with all actions that were committed in the merge. Finally, we computed the merge effort by subtracting the former multiset from the latter. The produced multiset contains just the lines of code added or removed during the merge operation, and the merge effort is the total number of actions in this multiset. For instance, a merge that combines two independent methods added in separate files would lead to zero merge effort, since the VCS would perform it automatically. Similarly, if these two independent methods are added to the same file, but in different regions, the merge effort would also be zero, since no additional actions would be needed to conciliate the branches. However, integrating a new feature implemented in parallel to an extensive refactoring would lead to a significant merge effort to adjust the feature to the new code organization imposed by the refactoring.

The following operations are performed over commits with two parents to extract the metrics. Since merges with more than two parents, called octopus, cannot have conflicts and manual edits by definition, these cases would necessarily have zero effort and are ignored. From a merge commit named $\small commit_{merge}$, its parent commits are obtained, and, from them, the commit in which they were derived, called $\small commit\_{base}$ is identified. Given this information, three diff operations are executed to obtain the actions performed in the merge commit. The first diff is performed between $\small commit_{base}$ and $\small commit\_{merge}$, thus obtaining the actions incorporated in the merge. The authors formally define merge actions as described in the formula: \textit{actions\textsubscript{merge}~= diff(commit\textsubscript{base}, commit\textsubscript{merge})}.


Then, the diff is executed between $\small commit_{base}$ and the two-parent commits of the merge commit to obtain the actions performed in the two branches. Therefore, the actions performed on branches 1 and 2 are formally defined, respectively, in the formulas: \textit{actions\textsubscript{branch1}~= diff(commit\textsubscript{base}, commit\textsubscript{parent1})} and \textit{actions\textsubscript{branch2}~= diff(commit\textsubscript{base}, commit\textsubscript{parent2})}.



These actions make it possible to identify the extra work implemented in a merge commit. For this, it is first necessary to identify the actions that were performed in one of the branches, which are calculated from the sum of the actions performed in branches 1 and 2, as shown in the formula: \textit{actions\textsubscript{branches} = actions\textsubscript{branches1} + actions\textsubscript{branches2}}.


Finally, to determine the extra work, we use the relative complement of the branch actions in the merge actions. The merge effort considered in the experiments of this work is used in an absolute way, applying the module operation in $\small actions_{extra}$, according to the formulas: \textit{action\textsubscript{extra}~= actions\textsubscript{merge} - actions\textsubscript{branches}} and \textit{effort~= $\mid$ actions\textsubscript{extra}$\mid$}.



To collect versioning information from projects, we used Python's pygit2\footnote{https://www.pygit2.org/} library, and to identify refactorings in commits, we used version 2.1 of the RefactoringMiner tool, released in March 2021, which can identify 62 different types of refactoring operations. In studies performed by RefactoringMiner's authors, the tool is reported to achieve 98\% of precision and 93\% of recall~\cite{tsantalis2016, tsantalis2013}, which makes it the current state-of-the-art tool for automated refactoring detection. When using RefactoringMiner, some commits took a long time to process, sometimes causing the process to hang. Therefore, we applied a timeout of 5 minutes. If RefactoringMiner does not finish processing a commit within 5 minutes, we terminate the process and skip to the next commit. We adopted 5 minutes in alignment with the literature~\cite{mahmoudi2019}. This timeout occurred in only 0.14\% of the commits of the analyzed projects.

\subsection{\textbf{Association Rules}} \label{sec:association_rules}

The extraction of association rules is an important task in data mining, whose goal is to find relationships among the attributes in a database~\cite{han_data_2011}. An association rule represents a relational pattern between data items in the application domain that happens with a specific frequency. The extraction of association rules is a technique in data mining that allows the identification of meaningful patterns from the data.

In an attempt to identify whether the occurrence of refactorings influences the merge effort, we use some attributes that quantify the refactorings involved in each branch of a merge commit, the total number of refactorings that took place in both branches, and the merge effort (code churn). These mined attributes are presented in Table~\ref{table:attributes}.

\begin{table}
\centering
\caption{Attributes considered in the analysis.}
\label{table:attributes}
\begin{tabular}{ll}
\toprule[1.0pt]
\textbf{Attribute} & \textbf{Description} \\
\midrule
$\small b1$ & number of refactorings in branch-1  \\ \\
$\small b2$ & number of refactorings in branch-2  \\ \\
$\small refactorings$ & \makecell[l]{number total of refactorings = $\small b1$ + $\small b2$} \\ \\
$\small effort$ & \makecell[l]{number of new lines included or excluded \\ in the merge commit (code churn)} \\
\bottomrule[1.0pt]
\end{tabular}
\end{table}

The method proposed in this work uses the concept of multidimensional association rules~\cite{han_data_2011}. Given a relation (or table) $\small D$, a multidimensional association rule $\small X \rightarrow Y$, defined on $\small D$, is an implication of the form: $\small X_{1}\wedge X_{2}\wedge\cdots\wedge X_{n} \rightarrow Y_{1}\wedge Y_{2}\wedge\cdots\wedge Y_{m}$, where $\small n\geq 1, m\geq 1,$ and $X_{i} (1\leq i\leq n)$ as well as $\small Y_{j} (1\leq j\leq m)$ are conditions defined in terms of the distinct attributes of $\small D$~\cite{witten_data_2016, han_data_2011}.

The rule $\small X\rightarrow Y$ indicates, with a certain degree of assurance, that the occurrence of the antecedent~$\small X$ implies the occurrence of the consequent~$\small Y$. The relevance of an association rule is evaluated by three main measures of interest: \textit{Support}, \textit{Confidence}, and \textit{Lift} \cite{witten_data_2016}. The \textit{Support} metric is defined by the percentage of instances in \textit{D} that satisfy the conditions of the antecedent and the conditions of the consequent. It is computed as follows: $\small Sup_{(X\rightarrow Y)}=T_{X\cup Y}/T$, where $\small T_{X\cup Y}$ represents the number of records in \textit{D} that satisfy the conditions in $\small X$ and the conditions in $\small Y$, and $\small T$ is the number of records in $\small D$. On the other hand, \textit{Confidence} represents the probability of occurrence of the consequent, given the occurrence of the antecedent. It is obtained in the following manner: $\small Conf_{(X\rightarrow Y)}=T_{X\cup Y}/T_{X}$, where $\small T_{X}$ represents the number of records in \textit{D} that satisfy the conditions of the antecedent $\small X$. \textit{Support} and \textit{Confidence} are used as a filter in the process of mining association rules, that is, only the rules characterized by having a minimum \textit{Support} and a minimum \textit{Confidence} (defined as input parameters) are extracted.

To better illustrate the calculation of the \textit{Support} and \textit{Confidence} measures, let $\small D$ be the relation shown in Table~\ref{table:basededadosexemplo}, which includes entries about refactorings and merge effort. The amounts of refactorings in each branch have been discretized into four ranges of values: zero~(``0''), units~(``u''), dozens~(``d''), and hundreds or more~(``$\geq100$''). Considering the rule R: \textit{b1~=~``u''}$~\wedge$ \textit{b2~=~``d''} $ \rightarrow$ \textit{effort~= ``true''}, we can find four records in $\small D$ that satisfy the three conditions of R, which are rows 1, 4, 6, and 8. Thus, $\small T_{(X \cup Y)}=4$ and, since $\small D$ has 8 entries ($\small T=8$), we can conclude that \textit{Sup(R)} = 50\%~(4/8). In this case, the \textit{Confidence} of the rule is 66.6\%~(4/6), since $\small T_{(X \cup Y)}=4$ and the conditions in the antecedent of the rule (\textit{b1~ =~``u''} $\wedge$ \textit{b2~=~``d''}) are satisfied in 6 entries ($\small T_{X}=6$), as we can see in rows 1, 3, 4, 6, 7, and 8.

Another measure of interest considered in this work is the \textit{Lift} of a rule $\small X\rightarrow Y$, which indicates how more frequently the conditions in $\small Y$ occurs given that the conditions in $\small X$ occur. \textit{Lift} is obtained by the quotient of the \textit{Confidence} of the rule and the \textit{Support} of its consequent, i.e., $\small Lift_{(X\rightarrow Y)}=Conf_{(X\rightarrow Y)}/Sup_{(Y)}$, where $\small Sup_{(Y)}$ represents the number of records in the relation that satisfy the conditions in $\small Y$. When $\small Lift=1 $ there is a conditional independence between $\small X$ and $\small Y$, that is, the antecedent does not interfere in the occurrence of the consequent. On the other hand, $\small Lift>1$ indicates a positive dependence between the antecedent and the consequent, meaning that the occurrence of $\small X$ increases the chances of the occurrence of $\small Y$. Conversely, when $\small Lift<1$ there is a negative dependence between the antecedent and the consequent, which indicates that the occurrence of $\small X$ decreases the chances of the occurrence of $\small Y$.

Taking into account the rule R used to exemplify the \textit{support} and \textit{confidence} measures, \textit{b1~=~``u''}$~\wedge~$\textit{b2~=~``d''} $ \rightarrow$ \textit{effort~= ``true''}, the \textit{Support} of the consequent (\textit{Sup(effort~=~``true'')}) of the rule in \textit{D} is equal to 50\%, that is, the percentage of entries that satisfy the condition \textit{effort = “true”}. Thus, the \textit{Lift} obtained for the rule R is 1.33, since \textit{Lift(R)~=~66.6/50~=~1.33}, where 66.6\% is the \textit{confidence} of the rule. In this case, the result indicates that, when there are few (units) of refactorings in branch 1 and some (dozens) in branch 2, the chances of having a merge effort increase by 33\%. In other words, we observe that the probability of the occurrence of \textit{effort~= ``true''} in $\small D$, which is 50\%, increases by a factor of 1.33 (becoming 66.6\%) given the occurrence of the antecedent \textit{b1~=~``u''}$~\wedge~$\textit{b2~=~``d''}.    

\begin{table}
\centering
\caption{Refactoring merge effort dataset sample.}
\label{table:basededadosexemplo}
\begin{tabular}{cccc}
\toprule[1.0pt]
\textbf{\#} & \textbf{b1} & \textbf{b2} & \textbf{effort} \\

\midrule
1 & ``u'' & ``d''         & ``true''  \\ 

2 & ``0'' & ``$\geq100$'' &  ``false'' \\ 

3 & ``u'' & ``d''         &  ``false'' \\ 

4 & ``u'' & ``d''         &  ``true'' \\ 

5 & ``u'' & ``u''         & ``false'' \\ 

6 &  ``u'' &``d''         & ``true'' \\ 

7 & ``u'' & ``d''         & ``false'' \\ 

8 & ``u'' & ``d''         & ``true'' \\ 
\bottomrule[1.0pt]
\end{tabular}
\end{table}

We have used \textit{Lift} values to avoid random implications, however \textit{Lift} is a symmetric metric. Consequently, as  $\small Lift_{(X\rightarrow Y)} = \small Lift_{(Y\rightarrow X)}$, we could not conclude whether X implies Y or vice-versa. Thus, we performed a confidence analysis to perceive the direction of the implication, especially when the difference between the confidence of both rules is significant. For example, when the confidence value of rule in the direction $\small X\rightarrow Y$ is significantly higher than the one in the direction $\small Y\rightarrow X$, we say that $\small X$ influences $\small Y$ and not the other way around.

We used the well-known Apriori \cite{agrawal_fast_1994} algorithm for the extraction of association rules, which is available in a Python library\footnote{http://http://rasbt.github.io/mlxtend/user\_guide/frequent\_patterns/apriori/}. We mined rules with minimum support of 0.05\%, given the large number of instances in our dataset, which represents a total of 20 merge commits.

\section{Results and Discussion} \label{sec:results}

In this section, we report and discuss the obtained results and present the answers to the research questions defined in Section~\ref{sec:methodology}. In Section~\ref{sec:resultsRQ1}, we analyze the influence of refactoring occurrences on the changes of having merge effort. Section~\ref{sec:resultsRQ2} focuses on the relationship between the number of refactorings and the merge effort. In Section~\ref{sec:resultsRQ3}, we discuss the association between the number of refactorings and the merge effort intensity. Finally, in Section~\ref{sec:resultsDiscussion} we present a discussion of the possible implications of our findings.

\subsection{RQ 1 - Does the occurrence of refactorings in the branches increase the chances of merge effort?} \label{sec:resultsRQ1}

\begin{figure}[!t]
\centering
\includegraphics[width=3.2in]{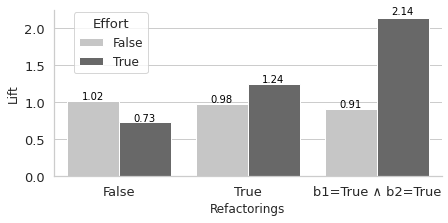}
\caption{Influence of refactorings on the occurrence of the merge effort.}
\label{fig:barchart_bin_bin}
\end{figure}

In an attempt to answer this research question, we initially discretized the dataset attributes to a binary domain, considering only a subset of attributes: number of refactorings in branch 1, number of refactorings in branch 2, the total number of refactorings, and merge effort. Therefore, we discretized each attribute as ``true'' or ``false'', indicating the occurrence (or not) of refactorings and merge effort. The graph in Figure~\ref{fig:barchart_bin_bin} summarizes the results. On the x-axis, we present three groups of association rules extracted from the discretized dataset: when no refactorings occurred in the branches (refactorings = ``false''); when some refactoring has been implemented in at least one of the branches (refactorings = ``true''); and when refactorings occurred in parallel in both branches (\textit{b1=``true''} $\land$ \textit{b2=``true''}). Light gray bars refer to merge commits that did not require any merge effort and dark gray bars refer to merge commits that required some effort. The y-axis represents the~\textit{Lift} value, calculated for each group of extracted rules, which indicates the strength of the relationship between the antecedent (i.e., the occurrence of refactorings) and the consequent (i.e., the occurrence of merge effort) of the rules.

At first, we decided to analyze the first two groups of rules, which only indicate the presence of refactorings, regardless of the branch in which they occurred. The ~\textit{Lift} values of the rules when there is no merge effort (light gray bars -- rules:~\textit{refactorings=``false''} $\rightarrow $~\textit{effort=``false''} and ~\textit{refactorings=``true''} $\rightarrow $~\textit{effort=``false''}) demonstrate that the presence or not of refactorings does not influence the non-occurrence of effort since these values are very close to 1. However, in the two rules where there is an effort (dark gray bars -- rules: ~\textit{refactorings=``false''} $\rightarrow $~\textit{effort=``true''} and ~\textit{ refactorings=``true''} $\rightarrow $~\textit{effort=``true''}), we can observe that \textbf{the occurrence of refactorings increases the chances of merge effort by 24\%} (~\textit{Lift = 1.24 }). Conversely, \textbf{having no refactorings decreases the chances of merge effort by 27\%} (~\textit{Lift = 0.73}).

Note that this discrepancy of almost no effect on the absence of effort and some effect on the occurrence of effort is a consequence of the frequency of merges with effort in our dataset. From a total of 40,248 valid merges, just 2,861 (7.1\% of the total) demanded some effort. Thus, a small fluctuation in the size of the light gray bars (from 1.02 to 0.98), which represent the bigger population of merges with no effort, leads to a big fluctuation in the size of the dark gray bars (from 0.73 to 1.24), which represent the small population of merges with some effort.

This increase of effort in the face of refactorings motivated us to check whether the occurrence of refactorings in parallel would be even more relevant, considering that some refactorings may be incompatible and lead to conflicts. Thus, we extracted the rule shown in the third group to assess whether the strength of this relationship would increase when refactorings were implemented simultaneously in both branches (rule: ~\textit{b1=``true''} $\land$ ~ \textit{b2=``true''} $\rightarrow $~\textit{effort=``true''}). This rule has a ~\textit{Lift} of 2.14, which indicates that \textbf{refactorings in both branches further increase the chances of merge effort by 114\%}.

\subsection{RQ 2 - Does the amount of refactorings in the branches increase the chances of merge effort?}
\label{sec:resultsRQ2}

To answer this research question, we kept the binary discretization of the effort attribute in the original dataset, that is, ``true'' or ``false''. However, the number of refactorings was discretized into four ranges of values: ``0'' (zero), ``u'' (units = 1 to 9), ``d'' (dozens = 10 to 99) and ``$\geq100$'' (hundreds or more). Thus, the antecedent of the rules extracted now can contain one of these four values. Considering the attribute number of refactorings (\textit{number\_of\_refactorings}) as the antecedent of the rules and the merge effort (\textit{effort}) as the consequence, we would have 8 possible rules. A rule in this case would have the following format: ~\textit{number\_of\_refactorings=``0'' / ``u'' / ``d'' / ``$\geq100$''} $\rightarrow$ effort=``true'' / ``false''. We chose not to subdivide the last range (``$\geq100$''') after analyzing that this range of values was the least frequent in all attributes of the dataset. In this way, we aim to minimize the imbalance of the intervals of values considered in this study.

\begin{figure}[!t]
\centering
\includegraphics[width=3.49in]{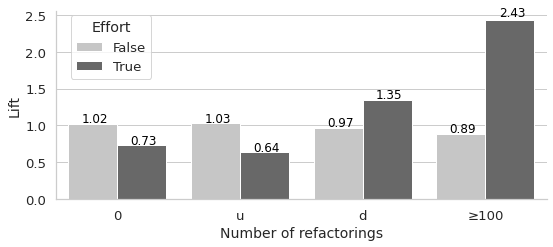}
\caption{Influence of the number of refactorings on the occurrence of the merge effort.}
\label{fig:barchart_mult_bin}
\end{figure}

The graph in Figure~\ref{fig:barchart_mult_bin} was constructed similarly to the graph in Figure~\ref{fig:barchart_bin_bin}. Nevertheless, the x-axis now represents the four ranges of values of the proposed discretization for the attribute~\textit{number\_of\_refactorings}: ``0'' (zero), ``u'' (units), ``d'' (dozens) and ``$\geq100$'' (hundreds or more). The \textit{Lift} values for the extracted rules demonstrate that the absence or the occurrence of few (units) refactorings tends to decrease the chances of merge effort. This behavior can be observed in the first two bars in dark gray color, with \textit{Lift} of 0.73 and 0.64. These values indicate that the complete absence of refactorings (``0'') decreases the chances of merge effort by 27\%, while refactorings in the units (``u'') range decrease by 36\% the changes of merge refactorings. As the number of refactorings increases, the possibility of a merge effort also increases. This statement can be verified in the dark gray bars with the number of refactorings in dozens (``d'') and hundreds or more (``$\geq100$''). \textbf{The occurrence of dozens of refactorings increases the chances of merge effort by 35\%} (\textit{Lift} = 1.35). \textbf{With a hundred or more refactorings, this increase is even greater: 143\%} (\textit{Lift} = 2.43). Analyzing the light gray bars, when there is no merge effort, we observe values very close to 1, showing no relevant relationship between the antecedent and the consequent of these association rules. Only in the last bar the \textit{Lift} value drops to 0.89, indicating that many refactorings decrease in 11\% the possibility of having no merge effort.

We observed some merge commits to assess whether the application of many refactorings was really necessary. In the project Netty, at the merge commit \texttt{81a6fb}\footnote{https://github.com/netty/netty/commit/81a6fb}, the developer applied three refactorings in the class \texttt{RuleBasedIpFilter}. These refactorings were applied to support code modifications of other refactorings in the classes \texttt{ChannelHandlerContext} and \texttt{InetSocketAddress} changed in the branches. In the merge commit, the developer applied two Extract Methods to create a new constructor, extracting code from the old constructor and the method \texttt{accept}, and a \texttt{Change Attribute Type} on the attribute rules. The method \texttt{accept} uses variables of the classes \texttt{ChannelHandlerContext} and \texttt{InetSocketAddress}. 
We observed the old constructor has a single line of code after these refactorings, 
and it calls the new constructor, adding a default boolean value as a parameter.  Although it is common in Object-Oriented languages such as Java to use method overhiding for implementing default parameters, here this strategy seems inappropriate because many code modifications were applied and developers may get confused whether the new constructor should be used or not.
This effort could be minimized if the developer applies an Extract Method to the old constructor that checks the boolean value, keeping the design simpler with just one constructor. Thus, the application of two Extract Methods was unnecessary in this case -- using just one Extract Method would lead to a better design and also decrease the merge effort. This example motivates us to suggest the creation of tools or guidelines for developers to avoid the application of unnecessary refactorings that can increase the merge effort. In this case, the tool could alert the developer about the problem to update the calls of the new constructor when it is necessary, or suggest other refactoring to minimize the effort.

\begin{figure}[!t]
\centering
\includegraphics[width=3.49in]{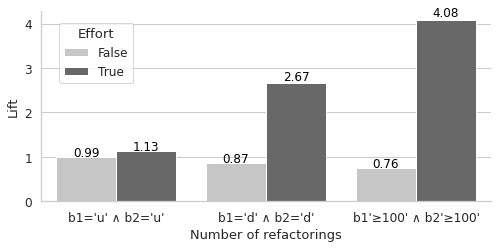}
\caption{Influence of the number of refactorings in parallel in both branches on the occurrence of merge effort.}
\label{fig:refactoring_mult_effort_bin_b1_b2}
\end{figure}

As we did in RQ1, we decided to evaluate whether increasing the number of refactorings in both merge branches increases the chances of merge effort. In this analysis, we do not just assess whether refactorings have taken place in the two branches, but we consider the number of refactorings according to the four discretization ranges. This would lead us to 32 possible rules, but we focused only on 6 of them with the same amount of refactorings in both branches. The graph in Figure~\ref{fig:refactoring_mult_effort_bin_b1_b2} presents on the x-axis three groups of rules whose antecedent considers the number of refactorings that co-occurred in both branches: the first one with refactorings in the range of the units, the second one in the range of dozens, and the third one in the range of hundreds or more. The occurrence of few refactorings in the two branches (\textit{b1=``u''} $\land$ \textit{b2=``u''}) increases by 13\% (\textit{Lift} = 1.13) the chances of having effort. This increase is even greater as the number of refactorings increases: \textbf{for dozens} (\textit{b1=``d''} $\land$ \textit{b2=``d''}) \textbf{of refactorings in both branches this percentage is 167\%} (\textit{Lift} = 2.67), \textbf{and with a hundred or more} (\textit{b1=``$\geq100$''} $\land$ \textit{b2=``$\geq100$''}) \textbf{refactorings it raises to impressing 308\%} (\textit{Lift} = 4.08). By analyzing the light gray bars, we verify the opposite behavior, although less intense. Note that the occurrence of many (``$\geq100$'') refactorings in both branches reduces by 24\% (\textit{Lift} = 0.76) the chances of having no effort.

\subsection{RQ 3 - Does the amount of refactorings in the branches increase the intensity of merge effort?} \label{sec:resultsRQ3}

To answer the third research question, we performed a new form of discretization of the original dataset. In addition to discretizing the antecedent into four ranges of values (``0'', ``u'', ``d'' and ``$\geq100$''), we did the same with the consequent, representing the merge effort. We discretized the consequent in this way to analyze to what extent the merge effort can be affected as we increase the number of refactorings in the branches of a merge commit.

\begin{figure}[!t]
\centering
\includegraphics[width=3.49in]{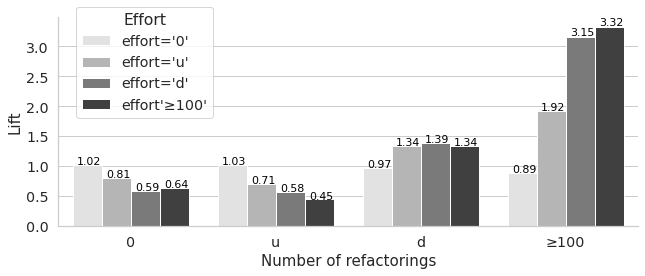}
\caption{Influence of the number of refactorings on the merge effort intensity.}
\label{fig:barchart_mult_mult}
\end{figure}

As mentioned, the focus of this research question is to evaluate the intensity of the merge effort given the number of refactorings that occurred in the branches of a merge commit. Thus, the graph in Figure~\ref{fig:barchart_mult_mult} presents four scales to represent the level of effort according to the proposed discretization for the consequent of the extracted association rules. These four scales are displayed in the graph legend in grayscale, where the lightest gray represents the absence of effort, and the darkest gray represents the greatest possible effort, i.e., hundreds or more (``$\geq100$'').

The results obtained for the first two blocks of rules, represented on the x-axis by zero~(``0'') and units~(``u''), demonstrate that fewer refactorings tend to generate fewer merge effort. This is particularly noticeable in the darkest gray bar, where, \textbf{for zero refactorings, the chances of having merge effort in the range of hundreds or more (``$\geq100$'') decrease by 36\%} (\textit{Lift} = 0.64), and \textbf{for units (``u''), this percentage is even higher, at 55\%} (\textit{Lift} = 0.45). Still in this part of the graph (refactorings = ``u''), it is possible to notice a decrease in the value of \textit{Lift}, reinforcing that the number of refactorings has a great influence on the merge effort. Analyzing the rules of the third group (refactorings = ``d''), we can notice that dozens of refactorings increase the chances of having merge effort in the intensity of ``u'', ``d'' and ``$\geq100$'' almost in the same proportion, by 34\% (\textit{Lift} = 1.34),  39\% (\textit{Lift} = 1.39),  and 34\% (\textit{Lift} = 1.34), respectivelly. This behavior is profoundly accentuated when we have many refactorings, as shown in the last rule block (refactorings = ``$\geq100$''). The \textit{Lift} values show that \textbf{many refactorings decrease the chances of having no merge effort by 11\%} (\textit{Lift} = 0.89), and \textbf{significantly increase the chances of having merge effort in units (``u''), in dozens (``d''), and in hundred or more (``$\geq100$'') ranges, presenting respectively the following percentages: 92\%} (\textit{Lift} = 1.92), \textbf{215\%} (\textit{Lift} = 3.15) \textbf{and 232\%} (\textit{Lift} = 3.32).

\begin{figure}
\centering
\includegraphics[width=3.49in]{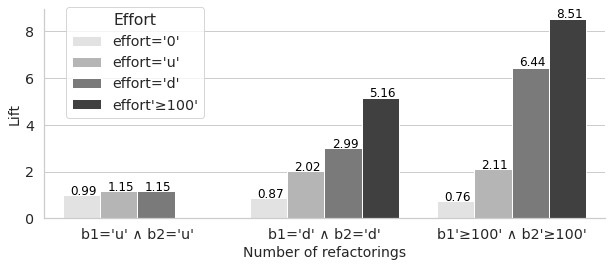}
\caption{Influence of the number of refactorings implemented simultaneously in both branches on the merge effort intensity.}\label{fig:refactoring_mult_effort_mult_b1_b2}
\end{figure}

In the same way as in RQ1 and RQ2, we analyzed the situation when refactorings co-occur in both branches. In this RQ, we aimed to assess whether, as the number of parallel refactorings increases in both branches, the intensity of the merge effort also increases. The results shown in the graph of Figure~\ref{fig:refactoring_mult_effort_mult_b1_b2} demonstrate that many refactorings occurring in parallel in both branches increase the chances of the merge effort being large. \textbf{For dozens of refactorings in the two branches} (\textit{b1=``d''} $\land$ \textit{b2=``d''}), for example, \textbf{this increase is quite noticeable, increasing by 102\%} (\textit{Lift } = 2.02) \textbf{the chances of the effort being in the units range, 199\%} (\textit{Lift} = 2.99) \textbf{of being in the dozens range, and 416\%} (\textit{Lift} = 5.16) \textbf{of being in the hundreds or more range}. The chances of no effort in this situation are reduced by 13\% (\textit{Lift} = 0.87). This behavior was even more evident \textbf{when hundreds or more} (\textit{b1=``$\geq100$''} $\land$ \textit{b2=``$\geq100$''}) \textbf{refactorings occurred in parallel in both branches}. In this case, \textbf{the chances of having effort in the units range increased by 111\%} (\textit{Lift} = 2.11), \textbf{of being in the dozens range by 544\%} (\textit{Lift} = 6.44), \textbf{and of being in the hundreds or more range by 751\%} (\textit{Lift} = 8.51). The chances of having no effort when there are hundreds or more refactorings in parallel in both branches are reduced even more, by 24\% (\textit{Lift} = 0.76). The first block of bars in the graph, when few refactorings occurred in parallel in both branches (\textit{b1=``u''} $\land$ \textit{b2=``u''}), also shows a slight increase in the chances of having effort in units and dozens, with 15\% in both cases. However, not enough merge cases were identified that meet the minimum support for the effort in the hundreds or more range in this situation. The results of these analyses demonstrate that the number of refactorings that occurred in parallel in both branches does not only influence the occurrence of merge effort but also its intensity.

\subsection{\textbf{Discussion}} \label{sec:resultsDiscussion}

In this section, we will discuss some implications based on the results of our experiments.

\textbf{Leveraging Collaborative Refactoring-Aware Changes}. From the point of view of systems development managers or even of developers (when there is no clearly defined manager role), our results may suggest some analysis on how to conduct the implementation of refactorings in a project: (i) analyze the feasibility of avoiding parallel branches to implement refactorings to minimize the chances of having high-intensity merge effort, (ii) avoid allocating extensive refactorings in parallel with other important changes, (iii) motivate developers to synchronize with their peers before extensive refactorings, (iv) evaluate the possibility of applying alternative techniques to the branches, such as the use of trunk-based development~\cite{trunkbased2022} with features toggles~\cite{rezvan2021}. The idea is to enable a more transparent collaborative development process so that new features are implemented directly on the mainline of development without creating branches. This strategy allows developers to be aware of parallel changes as soon as they are made, avoiding accumulating changes that only appear to other developers during the merge. Existing work~\cite{prutchi2022} already indicates a reduction of merge effort when trunk-based development is in place. Although merges of independent clones still occur, they are known to be less complex than merges of named branches, which are suppressed due to the use of feature flags.

\textbf{New Merge Strategies for Refactoring Changes}. From the tool builders' perspective, our results may indicate the need for new merge strategies, for example, first merging edits (without considering refactorings) and then replaying the refactorings. This approach may avoid the occurrence of merge semantic conflicts~\cite{mens2002}, as the refactorings would be merged into a branch that already contains all edits not classified as refactorings. This way of performing a merge involving refactorings can be inspired by the operation-based merging approach, as discussed in Mens~\cite{mens2002}. Moreover, the response to RQ3 may suggest creating tools that signal the most appropriate time to perform the merge, considering the number of refactorings implemented in both branches and the predicted merge effort level. In this way, developers could receive alerts to execute or even postpone the merge, considering a specific threshold of refactorings, thus potentially minimizing the merge effort.

\textbf{Improving Search-Based Refactoring Recommenders with Merge Effort Estimates}. Our findings in Sections~\ref{sec:resultsRQ1} and \ref{sec:resultsRQ2} also indicate there is room for proposing refactoring recommenders that better support developers working in parallel. Even though software maintenance is increasingly being parallelized, existing multi-objective refactoring approaches (e.g.\cite{morales2016,ouni2016,assuncao2021}) are not designed to take parallel changes into consideration. They are often defined in terms of objective functions (or criteria) capturing static~\cite{morales2016, ouni2016, assuncao2021} and dynamic~\cite{assuncao2021} quality attributes, feature dependencies~\cite{assuncao2021}, test coverage~\cite{ouni2016} and consistency with already made changes~\cite{ouni2016}.

This negligence may induce developers in selecting and performing refactorings in one of the branches that, although end up maximizing the satisfaction of all the aforementioned criteria, will either require or increase merge effort. If the other refactoring and non-refactoring changes being made in parallel are considered (even if partially), the recommender algorithms will search for alternative refactoring solutions that achieve the same objectives and are less prone to (higher) merge effort later. Another possible alternative is the recommender to simply suggest the refactoring is postponed until the possibly conflicting merge of the other branch is concluded. Finally, given our finding in Section~\ref{sec:resultsRQ3}, which indicated that shorter sequences of refactorings lead to lower merge effort, one using a multi-objective refactoring recommender might consider using the amount of refactorings as an additional objective function. Thus, the recommender would prioritize for shorter refactorings.

\textbf{Relating Refactoring Types and Merge Effort}. In addition, our results may encourage more specific studies on which types of refactorings tend to generate more merge effort. Still in this line, we intend to assess the incompatibility of refactorings in the branches of a merge, analyzing the following question: which types of refactorings (implemented in parallel in the branches) result in a greater chance of having a high-intensity merging effort?

\section{\textbf{Threats to Validity}} \label{sec:threats}

Some internal, external, construction, and conclusion threats may have influenced the results reported in this work, as presented in the following:

\textbf{Internal Validity}.
The association rule extraction technique helped understand the intensity of the relationships among the refactorings and merge effort attributes. However, a third variable not considered in our study may have affected both attributes. Moreover, as mentioned in Section~\ref{sec:refac_merge_effort}, when using RefactoringMiner to detect refactorings in a given commit, we employed a 5-minute timeout. We recorded every time RefactoringMiner took more than 5 minutes, and the process was terminated. Out of 428,109 commits analyzed with RefactoringMiner, only 615 (0.14\%) reached this timeout. This is a tiny percentage and does not pose a severe threat to the validity of our results.

\textbf{External Validity}. The process of building the project corpus may have disregarded projects relevant to our experiments. To mitigate this risk, we followed a systematic selection methodology, as described in section~\ref{sec:project_corpus}. In addition, our study focuses only on Java projects. However, as we were limited to open-source software systems and did not have access to closed-source enterprise Java projects, we cannot claim that our findings can be generalized to all Java software systems. Likewise, we cannot generalize our results to other programming languages. Thus, we note that the ability to generalize our findings is restricted to the characteristics of the selected projects. Our analysis, however, showed that there are patterns of behavior in the sample we chose, and we believe that they may also be present in other projects and languages. Nevertheless, Additional studies are needed. Additionally, although the RefactoringMiner 2.1 tool can identify 62 types of refactorings, our experiments considered only 33 types, which are the most frequently applied in other studies~\cite{bibiano2021}. We conjecture that a study evaluating a more extensive set of refactoring types would possibly find an even more substantial involvement of refactoring operations in increasing merge effort.

\textbf{Construct Validity}. Our methodology looked for refactorings that influence the merge effort. However, it is not easy to determine whether the refactoring was the only cause of the merge effort. This is because the refactorings implementation is often tangled with other types of changes~\cite{murphy2012}. Moreover, the merge effort calculation is based on the code churn (\textit{i.e.}, the number of lines added or removed) metric. Although this type of approach, considering the code churn, is widely used in software engineering to assess effort~\cite{olsson2018,coelho2021,prutchi2022}, qualitative evaluations may be necessary to confirm the obtained results. Additionally, we only considered merge commits with two parents during the dataset build process. In theory, a merge commit may have more than two parents, which is called \textit{octopus} in Git parlance. Nonetheless, in practice, an octopus merge commit rarely happens. In our dataset, for example, we found only four commits with three or more parents, representing only 0.007\% of the total merge commits (55,044). These four commits were found in only three projects: graal (1), jenkins (2), and spring-framework (1). Even in these rare cases, assessing the effort of \textit{octopus} merges is irrelevant, considering that, by definition, Git only allows octopus merges when there is no conflict and no manual edits.

\textbf{Conclusion Validity}. We first note that some association rules we identified have relatively little support. Therefore, some of them may have happened by chance. However, suppressing such rules would hinder rare but relevant rules. 

\section{\textbf{Related Work}} \label{sec:related}

We identified some works related to code refactorings and code merge operations~\cite{dig2007, laszlo2007, lebenich2017, mahmoudi2018, mahmoudi2019}. Most of them~\cite{dig2007, laszlo2007, lebenich2017, mahmoudi2018} focused on building, improving, or proposing merge tools that take into account the occurrence of refactorings in the branches. Only Mahmoudi et al.~\cite{mahmoudi2019} did more analytical work, evaluating the relationship between refactorings and merge conflicts. In this section, we will discuss the contributions and research gaps of these works in more detail.

Angyal et al. ~\cite{laszlo2007} extended the three-way merge algorithm to support renamings. The main idea is that the renamings should be detected before the merge and considered while reconciling the changes. Similarly, Dig et al.~\cite{dig2007} presented a software configuration management tool, called MolhadoRef, capable of identifying some types of refactorings before performing the code merge. They used the RefactoringCrawler tool, capable of identifying up to seven types of refactorings. MolhadoRef uses the operation-based approach, treating both refactorings and edits (other code changes not classified as refactorings) as change operations that are recorded and replayed. As edits and refactorings are mixed, they proposed inverting the refactorings, performing a textual merge, and replaying the refactoring operations. The main idea is to merge refactorings to eliminate merge errors and unnecessary merge conflicts. Leßenich et al.~\cite{lebenich2017} proposed improvements to their code merge tool (JDIME) to deal with some types of refactorings: renamings and code move. They used a heuristic method and evaluated their approach on real-world merge scenarios (48 projects and 4,878 merge scenarios). Mahmoudi and Nadi~\cite{mahmoudi2018} performed an empirical study to understand the nature of changes that phone vendors make versus modifications made in the original development of Android. The study focused on the simultaneous evolution of several versions of the Android operating system by investigating the overlap of different changes. Although the authors have not developed a tool, they have investigated and reported the most common types of refactorings that occur in practice and have looked at the possibility of having automated support for merging them. As a proxy case study, they analyzed the changes in the popular community-based variant of Android, LineageOS, and its corresponding Android versions. The experiments considered a small subset of refactorings (six different types), considered by the authors to be the most common.Nevertheless, while our work focused on assessing and quantifying the impact of refactorings on the merge effort, these workfocused on creating or improving merge tools that involve code refactorings. Their contributions have the potential to alleviate some of the problems discussed in our paper. 

Mahmoudi et al.~\cite{mahmoudi2019} carried out a large-scale empirical study, with about 3,000 open-source projects, to assess the relationship between 15 types of refactorings and the occurrence of merge conflicts. To analyze how often merge conflicts involve refactored code, they checked whether a code change that led to a conflict has involved a refactoring change. The term ``involved refactoring'' was used to describe a refactoring that occurred in an evolutionary change that overlaps the conflicting region. As a conflicting merge scenario can have multiple conflicting regions, if at least one of the conflicting regions in a conflicting merge scenario contained involved refactorings, they consider that the merge scenario has involved refactorings. As a result, they found out that 22\% of merge conflicts involve refactorings and that 11\% of conflicting regions have at least one refactoring involved. Their studies also presented that Extract Method refactorings are involved in more conflicts than their typical overall frequency, while most refactoring types are engaged in conflicts less frequently. The authors also conducted experiments to analyze whether conflicts involving refactorings are more difficult to resolve. For that, they used two metrics to define merge effort: the number of conflicting lines in the conflicting chunks and the number of commits with evolutionary changes that contain refactorings. The results show that conflicting regions involving refactorings tend to be larger (i.e., more complex) than those without refactorings. Furthermore, conflicting merge scenarios with involved refactorings include more evolutionary changes (i.e., changes that lead to conflict) than conflicting merge scenarios with no involved refactorings. 

As in our work, Mahmoudi et al.~\cite{mahmoudi2019} analyzed the consequences of refactorings on the merge operation. However, their experiments have considered the size of the conflicting chunks and not the effort required to perform the merge. It is essential to point out that an analysis purely based on individual chunks cannot perceive semantic conflicts, which can occur outside of chunks or even due to multiple simultaneous chunks. That is, non-conflicting merges (physical conflicts raised by the version control system) may generate syntactic or semantic code conflicts that potentially demand effort during the merge operation. By measuring the code churn required during the merge operation, we can quantify the total merge effort, even if is a consequence of syntactic or semantic conflicts. Furthermore, using the association rules extraction technique, our approach allowed us to quantify the intensity of the relationship between refactorings and merge effort. Our study also presented an initial evaluation on the place where the refactorings are implemented, that is, the branches. In this first moment, we focused on evaluating the parallelism of the refactorings in the two branches of a merge. The results showed that implementing refactorings in parallel in both branches generates a more significant merge effort. 


\section{Conclusion} \label{sec:conclusions}

This work presented a study that analyzes the relationship between the implementation of refactorings in the branches and the effort required to merge such branches. 
Our main results indicate that (i) the implementation of refactorings in the branches of a merge commit increases the chances of having merge effort; (ii) the number of refactorings implemented in the branches of a merge commit influences the occurrence of merge effort: the more refactorings, the greater the chance of effort; (iii) the number of refactorings implemented in the branches of a merge commit influences the intensity of the merge effort: the more refactorings, the greater the chance of the effort being of greater intensity; and (iv) refactorings co-occurring in both branches of the merge commit tend to increase even more the chances of there being effort, as well as the intensity of it.


As future work, we propose to carry out experiments to analyze which types of refactorings tend to demand more significant merge effort. In this paper, we found that refactorings in parallel could harm the merge process. However, not all parallel refactoring is necessarily harmful. By knowing which types of refactorings are more or less incompatible, we could improve the precision of our current result. In other words, instead of saying \textit{``do not do refactorings in parallel"}, we could say \textit{``do not do refactoring X in parallel with refactoring Y"}. Such a finding could motivate researchers to devise tools for alerting developers about incompatible refactorings, suggesting a branch synchronization before applying the refactoring, or even helping automatize the merge of incompatible refactorings.

We also propose to extend the experiment to a more extensive set of projects. A larger number of repositories will allow for a more focused analysis per project to identify the characteristics of the projects and their refactorings in the merge effort. In addition, we suggest using other attributes in the data mining process, such as: the size of commits, the existence of merge conflicts, and identifying whether the merge commit is associated with a pull request. Including the size of commits in the antecedent of the association rules can help understand this attribute's relationship with the merge effort. We expect larger commits to increase the chances of merge conflicts and, consequently, the effort for resolution. It is also interesting to analyze whether commits with refactorings tend to be larger on average than non-refactoring commits. The existence of conflicts can help to understand whether semantic conflicts result in less effort than the conflicts identified by Git at the time of the merge. Similarly, we can analyze whether merging a pull request branch into a project can lead to less merge effort since a code review process has been performed before the merge.

Furthermore, we intend to explore the compatibility of the types of refactorings that co-occur in both branches; that is, to discover those implemented in parallel in both branches that increase the chances of having effort and its intensity. Experiments can also be conducted for private software projects and in systems developed in other programming languages to analyze whether the behavior observed in this work repeats in projects with different characteristics.


\bibliographystyle{IEEEtran}
\bibliography{IEEEabrv,ICSE2023}

\begin{thebibliography}{10}
\providecommand{\url}[1]{#1}
\csname url@samestyle\endcsname
\providecommand{\newblock}{\relax}
\providecommand{\bibinfo}[2]{#2}
\providecommand{\BIBentrySTDinterwordspacing}{\spaceskip=0pt\relax}
\providecommand{\BIBentryALTinterwordstretchfactor}{4}
\providecommand{\BIBentryALTinterwordspacing}{\spaceskip=\fontdimen2\font plus
\BIBentryALTinterwordstretchfactor\fontdimen3\font minus
  \fontdimen4\font\relax}
\providecommand{\BIBforeignlanguage}[2]{{%
\expandafter\ifx\csname l@#1\endcsname\relax
\typeout{** WARNING: IEEEtran.bst: No hyphenation pattern has been}%
\typeout{** loaded for the language `#1'. Using the pattern for}%
\typeout{** the default language instead.}%
\else
\language=\csname l@#1\endcsname
\fi
#2}}
\providecommand{\BIBdecl}{\relax}
\BIBdecl

\bibitem{brun2011}
\BIBentryALTinterwordspacing
Y.~Brun, R.~Holmes, M.~D. Ernst, and D.~Notkin, ``Proactive detection of
  collaboration conflicts,'' in \emph{Proceedings of the 19th ACM SIGSOFT
  Symposium and the 13th European Conference on Foundations of Software
  Engineering}, ser. Proc. 19th ACM SIGSOFT Symp.and the 13th European Conf. on
  Foundations of Software Engineering (ESEC/FSE).\hskip 1em plus 0.5em minus
  0.4em\relax New York, NY, USA: Association for Computing Machinery, 2011, p.
  168–178. [Online]. Available: \url{https://doi.org/10.1145/2025113.2025139}
\BIBentrySTDinterwordspacing

\bibitem{kasi2013}
B.~K. Kasi and A.~Sarma, ``Cassandra: Proactive conflict minimization through
  optimized task scheduling,'' in \emph{Proceedings of the 2013 International
  Conference on Software Engineering}, ser. Proc. International Conference on
  Software Engineering (ICSE).\hskip 1em plus 0.5em minus 0.4em\relax IEEE
  Press, 2013, p. 732–741.

\bibitem{zimmermann2007}
\BIBentryALTinterwordspacing
T.~Zimmermann, ``Mining workspace updates in cvs,'' in \emph{Proceedings of the
  Fourth International Workshop on Mining Software Repositories}, ser. Proc.
  4th International Workshop on Mining Software Repositories (MSR).\hskip 1em
  plus 0.5em minus 0.4em\relax USA: IEEE Computer Society, 2007, p.~11.
  [Online]. Available: \url{https://doi.org/10.1109/MSR.2007.22}
\BIBentrySTDinterwordspacing

\bibitem{mens2002}
T.~Mens, ``A state-of-the-art survey on software merging,'' \emph{IEEE
  Transactions on Software Engineering}, vol.~28, no.~5, pp. 449--462, 2002.

\bibitem{Apel2011}
\BIBentryALTinterwordspacing
S.~Apel, J.~Liebig, B.~Brandl, C.~Lengauer, and C.~K\"{a}stner,
  ``Semistructured merge: Rethinking merge in revision control systems,'' in
  \emph{Proceedings of the 19th ACM SIGSOFT Symposium and the 13th European
  Conference on Foundations of Software Engineering}, ser. Proc. 19th ACM
  SIGSOFT Symp.and the 13th European Conf. on Foundations of Software
  Engineering.\hskip 1em plus 0.5em minus 0.4em\relax New York, NY, USA:
  Association for Computing Machinery, 2011, p. 190–200. [Online]. Available:
  \url{https://doi.org/10.1145/2025113.2025141}
\BIBentrySTDinterwordspacing

\bibitem{apiwattanapong2007}
\BIBentryALTinterwordspacing
T.~Apiwattanapong, A.~Orso, and M.~J. Harrold, ``Jdiff: A differencing
  technique and tool for object-oriented programs,'' \emph{Automated Software
  Eng.}, vol.~14, no.~1, p. 3–36, mar 2007. [Online]. Available:
  \url{https://doi.org/10.1007/s10515-006-0002-0}
\BIBentrySTDinterwordspacing

\bibitem{binkley1995}
\BIBentryALTinterwordspacing
D.~Binkley, S.~Horwitz, and T.~Reps, ``Program integration for languages with
  procedure calls,'' \emph{ACM Trans. Softw. Eng. Methodol.}, vol.~4, no.~1, p.
  3–35, jan 1995. [Online]. Available:
  \url{https://doi.org/10.1145/201055.201056}
\BIBentrySTDinterwordspacing

\bibitem{buffenbarger1995}
J.~Buffenbarger, \emph{Syntactic software merging}, J.~Estublier, Ed.\hskip 1em
  plus 0.5em minus 0.4em\relax Berlin, Heidelberg: Springer Berlin Heidelberg,
  1995.

\bibitem{hunt2002}
J.~Hunt and W.~Tichy, ``Extensible language-aware merging,'' in
  \emph{International Conference on Software Maintenance, 2002. Proceedings.},
  ser. International Conference on Software Maintenance, 2002, pp. 511--520.

\bibitem{shen2004}
H.~Shen and C.~Sun, ``A complete textual merging algorithm for software
  configuration management systems,'' in \emph{Proceedings of the 28th Annual
  International Computer Software and Applications Conference, 2004. COMPSAC
  2004.}, ser. Proceedings of the 28th Annual International Computer Software
  and Applications Conference (COMPSAC), 2004, pp. 293--298 vol.1.

\bibitem{shen2005}
------, ``Syntax-based reconciliation for asynchronous collaborative writing,''
  in \emph{2005 International Conference on Collaborative Computing:
  Networking, Applications and Worksharing}, ser. International Conference on
  Collaborative Computing: Networking, Applications and Worksharing, 2005, pp.
  10 pp.--.

\bibitem{westfechtel1991}
\BIBentryALTinterwordspacing
B.~Westfechtel, ``Structure-oriented merging of revisions of software
  documents,'' in \emph{Proceedings of the 3rd International Workshop on
  Software Configuration Management}, ser. Proc. 3rd International Workshop on
  Software Configuration Management (SCM).\hskip 1em plus 0.5em minus
  0.4em\relax New York, NY, USA: Association for Computing Machinery, 1991, p.
  68–79. [Online]. Available: \url{https://doi.org/10.1145/111062.111071}
\BIBentrySTDinterwordspacing

\bibitem{berzins1994}
\BIBentryALTinterwordspacing
V.~Berzins, ``Software merge: Semantics of combining changes to programs,''
  \emph{Journal ACM Transactions on Programming Languages and Systems},
  vol.~16, no.~6, p. 1875–1903, nov 1994. [Online]. Available:
  \url{https://doi.org/10.1145/197320.197403}
\BIBentrySTDinterwordspacing

\bibitem{jackson1994}
Jackson and Ladd, ``Semantic diff: a tool for summarizing the effects of
  modifications,'' in \emph{Proceedings 1994 International Conference on
  Software Maintenance}, ser. Proc. International Conference on Software
  Maintenance, 1994, pp. 243--252.

\bibitem{lebenich2015}
\BIBentryALTinterwordspacing
O.~LeBenich, S.~Apel, and C.~Lengauer, ``Balancing precision and performance in
  structured merge,'' \emph{Automated Software Eng.}, vol.~22, no.~3, p.
  367–397, sep 2015. [Online]. Available:
  \url{https://doi.org/10.1007/s10515-014-0151-5}
\BIBentrySTDinterwordspacing

\bibitem{apel2012}
S.~Apel, O.~LeBenich, and C.~Lengauer, ``Structured merge with auto-tuning:
  balancing precision and performance,'' in \emph{2012 Proceedings of the 27th
  IEEE/ACM International Conference on Automated Software Engineering}, ser.
  Proc.27th IEEE/ACM International Conference on Automated Software
  Engineering, 2012, pp. 120--129.

\bibitem{kim2012}
\BIBentryALTinterwordspacing
M.~Kim, T.~Zimmermann, and N.~Nagappan, ``A field study of refactoring
  challenges and benefits,'' in \emph{Proceedings of the ACM SIGSOFT 20th
  International Symposium on the Foundations of Software Engineering (FSE
  '12)}.\hskip 1em plus 0.5em minus 0.4em\relax New York, NY, USA: Association
  for Computing Machinery, 2012. [Online]. Available:
  \url{https://doi-org.ez24.periodicos.capes.gov.br/10.1145/2393596.2393655}
\BIBentrySTDinterwordspacing

\bibitem{fowler2018}
M.~Fowler, \emph{\BIBforeignlanguage{en}{Refactoring: Improving the Design of
  Existing Code}}, 2nd~ed.\hskip 1em plus 0.5em minus 0.4em\relax
  Addison-Wesley Professional, Nov. 2018.

\bibitem{dig2007}
D.~Dig, K.~Manzoor, R.~Johnson, and T.~N. Nguyen, ``Refactoring-aware
  configuration management for object-oriented programs,'' in \emph{29th
  International Conference on Software Engineering (ICSE'07)}, ser.
  International Conference on Software Engineering (ICSE'07), 2007, pp.
  427--436.

\bibitem{lebenich2017}
O.~LeBenich, S.~Apel, C.~Kästner, G.~Seibt, and J.~Siegmund, ``Renaming and
  shifted code in structured merging: Looking ahead for precision and
  performance,'' in \emph{2017 32nd IEEE/ACM International Conference on
  Automated Software Engineering (ASE)}, ser. 32nd IEEE/ACM International
  Conference on Automated Software Engineering (ASE), 2017, pp. 543--553.

\bibitem{mahmoudi2018}
\BIBentryALTinterwordspacing
M.~Mahmoudi and S.~Nadi, ``The android update problem: An empirical study,'' in
  \emph{Proceedings of the 15th International Conference on Mining Software
  Repositories}, ser. Proc. 15th International Conference on Mining Software
  Repositories (MSR).\hskip 1em plus 0.5em minus 0.4em\relax New York, NY, USA:
  Association for Computing Machinery, 2018, p. 220–230. [Online]. Available:
  \url{https://doi.org/10.1145/3196398.3196434}
\BIBentrySTDinterwordspacing

\bibitem{laszlo2007}
A.~László, L.~Lengyel, and H.~Charaf, ``Detecting renamings in three-way
  merging,'' \emph{Acta Polytechnica Hungarica}, vol. vol. 4, no. 4, 12 2007.

\bibitem{mahmoudi2019}
M.~Mahmoudi, S.~Nadi, and N.~Tsantalis, ``Are refactorings to blame? an
  empirical study of refactorings in merge conflicts,'' in \emph{2019 IEEE 26th
  International Conference on Software Analysis, Evolution and Reengineering
  (SANER)}, ser. IEEE 26th International Conference on Software Analysis,
  Evolution and Reengineering (SANER), 2019, pp. 151--162.

\bibitem{yamashita2013}
D.~I. Sjøberg, A.~Yamashita, B.~C. Anda, A.~Mockus, and T.~Dybå,
  ``Quantifying the effect of code smells on maintenance effort,'' \emph{IEEE
  Transactions on Software Engineering}, vol.~39, no.~8, pp. 1144--1156, 2013.

\bibitem{mcKee2017}
S.~McKee, N.~Nelson, A.~Sarma, and D.~Dig, ``Software practitioner perspectives
  on merge conflicts and resolutions,'' in \emph{2017 IEEE International
  Conference on Software Maintenance and Evolution (ICSME)}, 2017, pp.
  467--478.

\bibitem{brindescu2018}
\BIBentryALTinterwordspacing
C.~Brindescu, ``How do developers resolve merge conflicts? an investigation
  into the processes, tools, and improvements,'' in \emph{Proceedings of the
  2018 26th ACM Joint Meeting on European Software Engineering Conference and
  Symposium on the Foundations of Software Engineering}, ser. ESEC/FSE
  2018.\hskip 1em plus 0.5em minus 0.4em\relax New York, NY, USA: Association
  for Computing Machinery, 2018, p. 952–955. [Online]. Available:
  \url{https://doi.org/10.1145/3236024.3275430}
\BIBentrySTDinterwordspacing

\bibitem{kalliamvakou2014}
E.~Kalliamvakou, G.~Gousios, K.~Blincoe, L.~Singer, D.~M. German, and
  D.~Damian, ``The promises and perils of mining github.'' in \emph{Proceedings
  of the 11th Working Conference on Mining Software Repositories}, ser.
  Proc.11th Working Conference on Mining Software Repositories (MSR).\hskip 1em
  plus 0.5em minus 0.4em\relax Hyderabad, India: ACM, 2014, pp. 92--101.

\bibitem{borges2018}
H.~Borges and M.~T. Valente, ``What’s in a github star? understanding
  repository starring practices in a social coding platform.'' \emph{Journal of
  Systems and Software}, vol. 146, pp. 112–--129, 2018.

\bibitem{tiobe2022}
TIOBE, ``Tiobe index for august 2022,''
  \url{https://www.tiobe.com/tiobe-index/}, 2022, (accessed: 07.11.2022).

\bibitem{stackoverflow2021}
StackOverflow, ``Developer survey results,''
  \url{https://insights.stackoverflow.com/survey/2021/}, 2021, (accessed:
  07.11.2022).

\bibitem{tsantalis2020}
N.~Tsantalis, A.~Ketkar, and D.~Dig, ``Refactoringminer 2.0,'' \emph{IEEE
  Transactions on Software Engineering}, pp. 1--1, 2020.

\bibitem{tsantalis2018}
\BIBentryALTinterwordspacing
N.~Tsantalis, M.~Mansouri, L.~M. Eshkevari, D.~Mazinanian, and D.~Dig,
  ``Accurate and efficient refactoring detection in commit history,'' in
  \emph{Proceedings of the 40th International Conference on Software
  Engineering}, ser. Proc. IEEE/ACM 40th International Conference on Software
  Engineering (ICSE).\hskip 1em plus 0.5em minus 0.4em\relax New York, NY, USA:
  ACM, 2018, pp. 483--494. [Online]. Available:
  \url{http://doi.acm.org/10.1145/3180155.3180206}
\BIBentrySTDinterwordspacing

\bibitem{barnett1994}
V.~Barnett and T.~Lewis, \emph{\BIBforeignlanguage{en}{Outliers in Statistical
  Data}}, 3rd~ed.\hskip 1em plus 0.5em minus 0.4em\relax JSTOR, 1994.

\bibitem{bibiano2021}
A.~C. Bibiano, W.~K.~G. Assunção, D.~Coutinho, K.~Santos, V.~Soares,
  R.~Gheyi, A.~Garcia, B.~Fonseca, M.~Ribeiro, D.~Oliveira, C.~Barbosa, J.~L.
  Marques, and A.~Oliveira, ``Look ahead! revealing complete composite
  refactorings and their smelliness effects,'' in \emph{2021 IEEE International
  Conference on Software Maintenance and Evolution (ICSME)}, ser. IEEE
  International Conference on Software Maintenance and Evolution (ICSME), 2021,
  pp. 298--308.

\bibitem{prudencio2012}
\BIBentryALTinterwordspacing
Prudêncio, M.~João-Gustavo, W.~Leonardo, C.~Cláudia, and Rafael, ``To lock,
  or not to lock: That is the question.'' \emph{Journal of Systems and
  Software}, vol. 85(2), pp. 277–--289, 02 2012. [Online]. Available:
  \url{http://www.sciencedirect.com/science/article/pii/S0164121211001063}
\BIBentrySTDinterwordspacing

\bibitem{moura2018}
Moura, M.~Tayane, and Leonardo, ``Uma técnica para a quantificação do
  esforço de merge.'' in \emph{In Proceedings of the 6th Workshop on Software
  Visualization, Evolution and Maintenance}, ser. Proc. 6th Workshop on
  Software Visualization, Evolution and Maintenance.\hskip 1em plus 0.5em minus
  0.4em\relax São Carlos, SP, Brasil, 2018, 2018.

\bibitem{knuth1997}
Knuth and Donald-E, \emph{\BIBforeignlanguage{en}{The Art of Computer
  Programming}}, 3rd~ed.\hskip 1em plus 0.5em minus 0.4em\relax USA:
  Addison-Wesley Longman Publishing Co., 1997, vol.~2.

\bibitem{tsantalis2016}
D.~Silva, N.~Tsantalis, and M.-T. Valente, ``Why we refactor? confessions of
  github contributors.'' in \emph{In Proceedings of the 2016 24th ACM SIGSOFT -
  International Symposium on Foundations of Software Engineering - FSE}.\hskip
  1em plus 0.5em minus 0.4em\relax ACM Press, New York, New York, USA, 2015,
  pp. 858–--870.

\bibitem{tsantalis2013}
Tsantalis, G.~Nikolaos, S.~Victor, H.~Eleni, and Abram, ``A multidimensional
  empirical study on refactoring activity.'' \emph{In 23rd CASCON. IBM Corp.},
  pp. 132–--146, 2013.

\bibitem{han_data_2011}
J.~Han, M.~Kamber, and J.~Pei, \emph{\BIBforeignlanguage{en}{Data {Mining}:
  {Concepts} and {Techniques}: {Concepts} and {Techniques}}}, 3rd~ed.\hskip 1em
  plus 0.5em minus 0.4em\relax Elsevier, Jun. 2011.

\bibitem{witten_data_2016}
I.~H. Witten, E.~Frank, and M.~A. Hall, \emph{\BIBforeignlanguage{en}{Data
  {Mining}: {Practical} {Machine} {Learning} {Tools} and {Techniques}}},
  3rd~ed.\hskip 1em plus 0.5em minus 0.4em\relax Elsevier, Feb. 2016.

\bibitem{agrawal_fast_1994}
\BIBentryALTinterwordspacing
R.~Agrawal and R.~Srikant, ``Fast {Algorithms} for {Mining} {Association}
  {Rules} in {Large} {Databases},'' in \emph{Proceedings of the 20th
  {International} {Conference} on {Very} {Large} {Data} {Bases}}, ser. Proc.
  20th International Conference on Very Large Data Bases (VLDB).\hskip 1em plus
  0.5em minus 0.4em\relax San Francisco, CA, USA: Morgan Kaufmann Publishers
  Inc., 1994, pp. 487--499. [Online]. Available:
  \url{http://dl.acm.org/citation.cfm?id=645920.672836}
\BIBentrySTDinterwordspacing

\bibitem{trunkbased2022}
P.~Hammant, ``What is trunk-based development?''
  \url{https://paulhammant.com/2013/04/05/what-is-trunk-based-development/},
  2013, (accessed: 07.11.2022).

\bibitem{rezvan2021}
\BIBentryALTinterwordspacing
R.~Mahdavi{-}Hezaveh, J.~Dremann, and L.~A. Williams, ``Feature toggle driven
  development: Practices usedby practitioners,'' \emph{CoRR}, vol.
  abs/1907.06157, 2019. [Online]. Available:
  \url{http://arxiv.org/abs/1907.06157}
\BIBentrySTDinterwordspacing

\bibitem{prutchi2022}
\BIBentryALTinterwordspacing
E.~S. Prutchi, H.~de~S.~Campos~Junior, and L.~G.~P. Murta, ``How the adoption
  of feature toggles correlates with branch merges and defects in open-source
  projects?'' \emph{Software: Practice and Experience}, vol.~52, no.~2, pp.
  506--536, 2022. [Online]. Available:
  \url{https://onlinelibrary.wiley.com/doi/abs/10.1002/spe.3034}
\BIBentrySTDinterwordspacing

\bibitem{morales2016}
R.~Morales, A.~Sabane, P.~Musavi, F.~Khomh, F.~Chicano, and G.~Antoniol,
  ``Finding the best compromise between design quality and testing effort
  during refactoring,'' in \emph{2016 IEEE 23rd International Conference on
  Software Analysis, Evolution, and Reengineering (SANER)}, vol.~1, 2016, pp.
  24--35.

\bibitem{ouni2016}
\BIBentryALTinterwordspacing
A.~Ouni, M.~Kessentini, H.~Sahraoui, K.~Inoue, and K.~Deb, ``Multi-criteria
  code refactoring using search-based software engineering: An industrial case
  study,'' \emph{ACM Trans. Softw. Eng. Methodol.}, vol.~25, no.~3, jun 2016.
  [Online]. Available: \url{https://doi.org/10.1145/2932631}
\BIBentrySTDinterwordspacing

\bibitem{assuncao2021}
W.~K.~G. Assunção, T.~E. Colanzi, L.~Carvalho, J.~A. Pereira, A.~Garcia,
  M.~J. de~Lima, and C.~Lucena, ``A multi-criteria strategy for redesigning
  legacy features as microservices: An industrial case study,'' in \emph{2021
  IEEE International Conference on Software Analysis, Evolution and
  Reengineering (SANER)}, 2021, pp. 377--387.

\bibitem{murphy2012}
E.~Murphy-Hill, C.~Parnin, and A.~P. Black, ``How we refactor, and how we know
  it,'' \emph{IEEE Transactions on Software Engineering}, vol.~38, no.~1, pp.
  5--18, 2012.

\bibitem{olsson2018}
T.~Olsson, M.~Ericsson, and A.~Wingkvist, ``Poster: Using repository data for
  driving software architecture,'' in \emph{2018 IEEE/ACM 40th International
  Conference on Software Engineering: Companion (ICSE-Companion)}, 2018, pp.
  197--198.

\bibitem{coelho2021}
\BIBentryALTinterwordspacing
F.~Coelho, N.~Tsantalis, T.~Massoni, and E.~L.~G. Alves, ``An empirical study
  on refactoring-inducing pull requests,'' in \emph{Proceedings of the 15th ACM
  / IEEE International Symposium on Empirical Software Engineering and
  Measurement (ESEM)}, ser. ESEM '21.\hskip 1em plus 0.5em minus 0.4em\relax
  New York, NY, USA: Association for Computing Machinery, 2021. [Online].
  Available: \url{https://doi.org/10.1145/3475716.3475785}
\BIBentrySTDinterwordspacing

\end{thebibliography}

\end{document}